\documentclass[11pt]{article}%
\usepackage{amsfonts}
\usepackage{amsmath}
\usepackage{amssymb}
\usepackage{graphicx}%
\providecommand{\U}[1]{\protect\rule{.1in}{.1in}}
\newtheorem{theorem}{Theorem}

\newtheorem{proposition}[theorem]{Proposition}

\begin{document}

\title{The Unique Pure Gaussian State Determined by the Partial Saturation of the
Uncertainty Relations of a Mixed Gaussian State}
\author{Maurice A. de Gosson\thanks{maurice.de.gosson@univie.ac.at}\\University of Vienna\\Faculty of Mathematics-NuHAG\\Nordbergstr. 15, 1090 Vienna}
\maketitle

\begin{abstract}
Let $\rho$ the density matrix of a mixed Gaussian state. Assuming that one of
the Robertson--Schr\"{o}dinger uncertainty inequalities is saturated by $\rho
$, \emph{e.g.} $(\Delta^{\rho}X_{1})^{2}(\Delta^{\rho}P_{1})^{2}=\Delta^{\rho
}(X_{1},P_{1})^{2}+\tfrac{1}{4}\hbar^{2}$, we show that there exists a unique
pure Gaussian state whose Wigner distribution is dominated by that of $\rho$
and having the same variances and covariance $\Delta^{\rho}X_{1},\Delta^{\rho
}P_{1}$, and $\Delta^{\rho}(X_{1},P_{1})$ as $\rho$. This property can be
viewed as an analytic version of Gromov's non-squeezing theorem in the linear
case, which implies that the intersection of a symplectic ball by a single
plane of conjugate coordinates determines the radius of this ball.

\end{abstract}

\section{Statement of Results}

Consider a mixed quantum state, identified with its density matrix $\rho$, and
let
\[
W\rho(x,p)=\left(  \tfrac{1}{2\pi\hbar}\right)  ^{n}\int_{\mathbb{R}^{n}%
}e^{-\frac{i}{\hbar}py}\langle x+\tfrac{1}{2}y|\rho|x-\tfrac{1}{2}y\rangle dy
\]
be its Wigner function. Viewing the latter as a quasi-probability distribution
we assume that it satisfies
\begin{equation}
\int\nolimits_{\mathbb{R}^{2n}}(1+|z|^{2})|W\rho(z)|dz<\infty\label{cond25}%
\end{equation}
(we have set $z=(x,p)$) so that the first and second moments of $W\rho$ exist.
The covariance matrix (CM) of $\rho$ is defined by
\begin{equation}
\Sigma_{\rho}=\int_{\mathbb{R}^{2n}}(z-\langle z\rangle)(z-\langle
z\rangle)^{T}W\rho(z)dz\label{sigmaz1}%
\end{equation}
where $\langle z\rangle=\operatorname*{Tr}(z\rho)$ is the mean value vector.
It will be convenient to write
\begin{equation}
\Sigma_{\rho}=%
\begin{pmatrix}
\Delta^{\rho}(X,X) & \Delta^{\rho}(X,P)\\
\Delta^{\rho}(X,P) & \Delta^{\rho}(P,P)
\end{pmatrix}
\label{sigmablock1}%
\end{equation}
where $\Delta^{\rho}(X,X)=(\Delta^{\rho}(X_{i},X_{j}))_{1\leq i,j\leq n}$,
$\Delta^{\rho}(X,P)=(\Delta^{\rho}(X_{i},P_{j}))_{1\leq i,j\leq n}$,
$\Delta^{\rho}(P,P)=(\Delta^{\rho}(P_{i},P_{j}))_{1\leq i,j\leq n}$ where
$\Delta^{\rho}(X_{i},X_{j})$, etc., are the covariances. Setting
$(\Delta^{\rho}X_{j})^{2}=\Delta^{\rho}(X_{j},X_{j})$ and $(\Delta^{\rho}%
P_{j})^{2}=\Delta^{\rho}(P_{j},P_{j})$ the Robertson--Schr\"{o}dinger (RS)
inequalities%
\begin{equation}
(\Delta^{\rho}X_{j})^{2}(\Delta^{\rho}P_{j})^{2}\geq\Delta^{\rho}(X_{j}%
,P_{j})^{2}+\tfrac{1}{4}\hbar^{2}\label{RS}%
\end{equation}
hold for $j=1,...,n$. We will say that these inequalities are \emph{partially
saturated} if at least one (but not all) of them are equalities.

We now assume that the state $\rho$ is Gaussian; this means that%
\begin{equation}
W\rho(z)=\left(  \frac{1}{2\pi}\right)  ^{n/2}(\det\Sigma)^{-1/2}\exp\left[
-\frac{1}{2}(z-\langle z\rangle)^{T}\Sigma^{-1}(z-\langle z\rangle)\right]
\label{rogauss}%
\end{equation}
(thus $\Sigma$ is assumed to be invertible, but this is no restriction: see
section \ref{sec2}). We will show that:

\begin{theorem}
\label{th1}Suppose that anyone of the (RS) inequalities (\ref{RS}) is
saturated, for instance
\begin{equation}
(\Delta^{\rho}X_{1})^{2}(\Delta^{\rho}P_{1})^{2}=\Delta^{\rho}(X_{1}%
,P_{1})^{2}+\tfrac{1}{4}\hbar^{2}. \label{1}%
\end{equation}
Then:

(i) There exists a unique pure Gaussian state $\psi$ such that%
\[
\Delta^{\psi}X_{1}=\Delta^{\rho}X_{1}\text{ , }\Delta^{\psi}P_{1}=\Delta
^{\rho}P_{1}\text{ , }\Delta^{\psi}(X_{1},P_{1})=\Delta^{\rho}(X_{1},P_{1})
\]

(ii) That state $\psi$ is the only Gaussian whose Wigner function satisfies
the inequality.%
\begin{equation}
W\psi(x,p)\leq W\rho(x,p). \label{ouipsi1}%
\end{equation}

\end{theorem}

We emphasize that no assumptions are made on the other variances or
covariances of the state $\rho$: they may for instance unknown.

One of the crucial steps in the proof is of a topological nature, and is
deeply related to Gromov's \cite{Gromov} symplectic non-squeezing theorem
(nicknamed by V.I. Arnol'd \cite{arnold} the \textquotedblleft principle of
the symplectic camel\textquotedblright; see \cite{FP} for a discussion of that
terminology). Let us now call \textquotedblleft symplectic
ball\textquotedblright\ the image $S(B_{R})$ of a phase space ball $B_{R}$
with radius $R$ by a linear (or affine) symplectic transformation $S$; the
number $R$ is the radius of the symplectic ball. Since symplectic
transformations are volume preserving, the volume of $S(B_{R})$ is the same as
that of the ball $B_{R}$. More surprising is the following rather unexpected
property: the section of the ellipsoid $S(B_{R})$ by any plane of conjugate
coordinates $x_{j},p_{j}$ passing through its center is always $\pi R^{2}$.
Thus, a single \textquotedblleft tomography\textquotedblright\ of a symplectic
ball unambiguously determines its radius! This is counterintuitive, because
one would expect that the area of the section of an ellipsoid by different
planes yields different results. It turns out that this property (of which we
give a proof in the Appendix) is equivalent to the linear version of Gromov's
theorem (see \cite{physreps} for a review of Gromov's theory; also
\cite{Polter} for various developments). Its relation with Theorem \ref{th1}
comes from the fact that the Wigner transform of a Gaussian is a Gaussian of a
very special type: its covariance ellipsoid is a symplectic ball, and the
state is thus entirely determined by a section of this ellipsoid by a single
plane of conjugate variables.

This paper has two forerunners \cite{physlett} and \cite{Bullsci} where we
showed that it was possible to associate a unique pure Gaussian state to a
mixed state satisfying a certain topological condition related to the
uncertainty principle, see (\ref{opt2}) in section \ref{sec2}.\bigskip

\noindent\textbf{Notation.} We will use the following notation and
terminology: The generic variable of phase space $\mathbb{R}^{2n}$ is
$z=(x,p)$ with $x=(x_{1},...,x_{n})$, $p=(p_{1},...,p_{n})$; in calculations
$x,p,z$ will always be viewed as column vectors. $\operatorname*{Sp}%
(2n,\mathbb{R})$ is the symplectic group; it consists of all real $2n\times2n$
matrices $S$ such that $S^{T}JS=J$ (or, equivalently, $SJS^{T}=J$) where $J=%
\begin{pmatrix}
0 & I\\
-I & 0
\end{pmatrix}
$. We use Dirac's notation $\hbar$ for $h/2\pi$ ($h$ Planck's
constant).\bigskip

\noindent\textbf{Acknowledgements}. The present work has been supported by the
Austrian Research Agency FWF (Projektnummer P20442-N13).

\section{Geometry and Uncertainty\label{sec2}}

In his seminal paper \textquotedblleft Geometry and
Uncertainty\textquotedblright\ \cite{Narcow} F. Narcowich (also see
\cite{Narconnell,Yuen}) proves, among other things, the following deep result:
a real symmetric matrix $\Sigma$ is the covariance matrix of a quantum state
if and only if
\begin{equation}
\Sigma+\frac{i\hbar}{2}J\text{ \emph{is positive semi-definite}} \label{opt1}%
\end{equation}
(abbreviated as $\Sigma+\frac{i\hbar}{2}J\geq0$). Condition (\ref{opt1})
implies (\cite{Narcow}, Lemma 2.3) that $\Sigma$ is fact definite positive,
and hence invertible. It will be convenient to use the auxiliary matrix
\begin{equation}
M=\frac{\hbar}{2}\Sigma^{-1} \label{M}%
\end{equation}
in which case (\ref{opt1}) becomes
\begin{equation}
M^{-1}+iJ\geq0. \label{opt1bis}%
\end{equation}
Narcowich's result seems to have become \textquotedblleft folk
wisdom\textquotedblright\ (and is therefore not very much cited); its proof
uses arguments from hard harmonic analysis (a symplectic version of Bochner's
theorem on positive measures \cite{Kastler,LouMiracle1,LouMiracle2}) and is
often stated (most of the time without proof) in the quantum-optical
literature; see for instance the references in \cite{adesso}. A
\textit{caveat:} as we have shown in \cite{GoLu0} there exist self-adjoint
operators with trace one whose covariance matrix satisfies condition
(\ref{opt1}), but which are not non-negative, and hence do not represent a
quantum state. However, given a covariance matrix $\Sigma$ satisfying
(\ref{opt1}) one can always construct a quantum state with covariance matrix
$\Sigma$, namely the Gaussian state $\rho$ whose Wigner transform is%
\[
W\rho(z)=\left(  \frac{1}{\pi\hbar}\right)  ^{n}(\det M)\exp\left(  -\frac
{1}{\hbar}z^{T}Mz\right)  .
\]

In \cite{Birk,mdg,FP,physreps,gostat} we have shown that condition
(\ref{opt1}) is equivalent to a topological statement involving the symplectic
capacity of the covariance ellipsoid%
\begin{equation}
\Omega=\{z:\tfrac{1}{2}z^{T}\Sigma^{-1}z\leq1\}=\{z:z^{T}Mz\leq\hbar\}.
\label{covell}%
\end{equation}
In fact, the algebraic condition $M^{-1}+iJ\geq0.$ is equivalent to the
following property:%
\begin{equation}
\text{\emph{The symplectic capacity} }c(\Omega)\text{ \emph{of the covariance}
\emph{ellipsoid} \emph{is }}\geq\tfrac{1}{2}h. \label{opt2}%
\end{equation}
The symplectic capacity $c(\Omega)$ is defined as follows: if $n=1$ it is just
the area of the ellipse $\Omega$; in higher dimensions it is the supremum of
all numbers $\pi R^{2}$ such that a symplectic ball $S(B_{R})$ is contained in
$\Omega$. In \cite{hileyfest} we have called a symplectic ball with radius
$\sqrt{\hbar}$ a \textquotedblleft quantum blob\textquotedblright.

Both conditions (\ref{opt1}), (\ref{opt2}) are, for a given quantum system,
equivalent to the Robertson--Schr\"{o}dinger (RS) inequalities%
\begin{equation}
(\Delta^{\rho}X_{j})^{2}(\Delta^{\rho}P_{j})^{2}\geq\Delta^{\rho}(X_{j}%
,P_{j})^{2}+\tfrac{1}{4}\hbar^{2}\text{ , }1\leq j\leq n. \label{RS1}%
\end{equation}
Using either (\ref{opt1}) or (\ref{opt2}) it is easy to show that the RS
inequalities are covariant (\textit{i.e.} they retain they form) under linear
or affine symplectic transformations. If we set $(X^{\prime},P^{\prime
})=S(X,P)$ where $S$ is a symplectic matrix, then
\begin{equation}
(\Delta^{\rho}X_{j}^{\prime})^{2}(\Delta^{\rho}P_{j}^{\prime})^{2}\geq
\Delta^{\rho}(X_{j}^{\prime},P_{j}^{\prime})^{2}+\tfrac{1}{4}\hbar^{2}\text{ ,
}1\leq j\leq n \label{RS1prime}%
\end{equation}
since the new covariance matrix is $\Sigma^{\prime}=S\Sigma S^{T}$ and%
\[
\Sigma^{\prime}+\frac{i\hbar}{2}J=S\left(  \Sigma+\frac{i\hbar}{2}J\right)
S^{T}\geq0
\]
because $SJS^{T}J=J$ as $S$ is symplectic.

The key to the proof of the equivalences (\ref{opt1})$\Longleftrightarrow
$(\ref{opt2})$\Longleftrightarrow$(\ref{RS1}) is the following well-known
symplectic diagonalization result (Williamson's \cite{wi36} theorem, of which
\cite{simon} gives an elementary proof). Let $M$ be a (real) symmetric
positive-definite matrix of size $2n$. There exists $S\in\operatorname*{Sp}%
(2n,\mathbb{R})$ and positive numbers $\lambda_{1},...,\lambda_{n}$ such that%
\begin{equation}
S^{T}MS=%
\begin{pmatrix}
\Lambda & 0\\
0 & \Lambda
\end{pmatrix}
\text{ \ , \ }\Lambda=\operatorname*{diag}(\lambda_{1},...,\lambda
_{n}).\label{will}%
\end{equation}
The numbers $\lambda_{1},...,\lambda_{n}$ are called the \emph{symplectic
eigenvalues} of $M$\ (or sometimes \emph{Williamson invariants}); they are
written in decreasing order: $\lambda_{1}\geq\cdot\cdot\cdot\geq\lambda_{n}$
and the array
\begin{equation}
\operatorname*{Spec}\nolimits_{\sigma}(M)=(\lambda_{1},...,\lambda
_{n})\label{spec}%
\end{equation}
is the \emph{symplectic spectrum} of $M$. The symplectic eigenvalues are
calculated as follows: consider the product $JM$ of $M$ by the standard
symplectic matrix and let $M^{1/2}$ be the positive square root of $M$; the
matrices $M^{1/2}JM^{1/2}$ and $JM$ are equivalent, and hence have the same
eigenvalues. Since $M^{1/2}JM^{1/2}$ is antisymmetric (because $J$ is), these
eigenvalues occur in pairs $(i\lambda_{j},\pm i\lambda_{j})$, $\lambda_{j}>0$,
and the $\lambda_{j}$ are precisely the symplectic eigenvalues of $M$.

The symplectic spectrum has the following straightforward properties:
if\ $c>0$%
\begin{equation}
\operatorname*{Spec}\nolimits_{\sigma}(cM)=(c\lambda_{1},...,c\lambda_{n})
\label{spec1}%
\end{equation}
and%
\begin{equation}
\operatorname*{Spec}\nolimits_{\sigma}(M^{-1})=(\lambda_{n}^{-1}%
,...,\lambda_{1}^{-1}). \label{spec2}%
\end{equation}
The relation between symplectic spectrum and symplectic capacity is essential:
let $M$ be a symmetric positive definite matrix and consider the phase space
ellipsoid $\Omega:z^{T}Mz\leq\hbar$. We have%
\begin{equation}
c(\Omega)=\frac{\pi\hbar}{\lambda_{1}}\text{ \ if \ }\operatorname*{Spec}%
\nolimits_{\sigma}(M)=(\lambda_{1},...,\lambda_{n}). \label{CM}%
\end{equation}

\section{The Wigner function of a Gaussian}

We will consider (normalized) Gaussians of the type
\begin{equation}
\psi_{X,Y}(x)=(\pi\hbar)^{-n/4}(\det X)^{1/4}e^{-\frac{1}{2\hbar}x^{T}(X+iY)x}
\label{Gauss1}%
\end{equation}
where $X$ and $Y$ are real symmetric $n\times n$ matrices, $X$ positive definite.

We will need the two following properties of the Wigner function
\begin{equation}
W\psi(x,p)=\left(  \tfrac{1}{2\pi\hbar}\right)  ^{n}\int_{\mathbb{R}^{n}%
}e^{-\frac{i}{\hbar}py}\psi(x+\tfrac{1}{2}y)\psi^{\ast}(x-\tfrac{1}{2}y)dy
\label{wig1}%
\end{equation}
(for a complete review of the Wigner formalism we refer to Littlejohn
\cite{Littlejohn} and de Gosson \cite{Birk,Birkbis}):

\noindent\textbf{Symplectic covariance}. Let $S\in\operatorname*{Sp}%
(2n,\mathbb{R})$ and $\widehat{S}$ be any of the two metaplectic operators
associated with $S$. Then
\begin{equation}
W(\widehat{S}\psi)(z)=W\psi(S^{-1}z). \label{metaco2}%
\end{equation}
Recall that the metaplectic operators are defined as follows: the symplectic
group $\operatorname*{Sp}(2n,\mathbb{R})$ has a covering group of order two,
the metaplectic group $\operatorname*{Mp}(2n,\mathbb{R})$. That group consists
of unitary operators on $L^{2}(\mathbb{R}^{n})$, and is generated by the
following elementary unitary transformations:

\begin{itemize}
\item The modified \textit{Fourier transform}%
\[
F\psi(x)=\left(  \tfrac{1}{2\pi i}\right)  ^{n/2}\int_{\mathbb{R}^{n}%
}e^{-\frac{i}{\hbar}x^{T}x^{\prime}}\psi(x^{\prime})dx^{\prime}%
\]
(with the convention $i^{1/2}=e^{i\pi/4}$);

\item \textquotedblleft\textit{Chirps}\textquotedblright\
\[
V_{P}\psi(x)=\exp\left(  -\frac{i}{2\hbar}x^{T}Px\right)  \text{ \ ,
\ }P=P^{T};
\]

\item \textit{Dilations}%
\[
M_{L}\psi(x)=\sqrt{\det L}\psi(Lx)\text{ \ , \ }\det L\neq0
\]
where $\sqrt{\det L}=i^{m}\sqrt{|\det L|}$, $m=0$ or $2$ if $\det L>0$ and
$m=1$ or $3$ if $\det L<0.$
\end{itemize}

\noindent For a detailed study of the metaplectic group $\operatorname*{Mp}%
(2n,\mathbb{R})$ see \cite{Birk}.

\noindent\textbf{Transformation of Gaussians}. The Wigner transform of a
Gaussian is itself a Gaussian. In fact%
\begin{equation}
W\psi_{X,Y}(z)=(\pi\hbar)^{-n}e^{-\frac{1}{\hbar}z^{T}Gz} \label{wg}%
\end{equation}
where $G$ is the real $2n\times2n$ matrix
\begin{equation}
G=%
\begin{pmatrix}
X+YX^{-1}Y & YX^{-1}\\
X^{-1}Y & X^{-1}%
\end{pmatrix}
. \label{g}%
\end{equation}

It turns out that the matrix $G$ defined in the formula above is both
positive-definite and symplectic: we have $G=S^{T}S$ where
\begin{equation}
S=%
\begin{pmatrix}
X^{1/2} & 0\\
X^{-1/2}Y & X^{-1/2}%
\end{pmatrix}
\label{ass}%
\end{equation}
and $S$ is obviously in $\operatorname*{Sp}(2n,\mathbb{R})$. The argument can
be reversed: given a positive-definite symplectic matrix $G$, we can always
find $S\in\operatorname*{Sp}(2n,\mathbb{R})$ such that $G=S^{T}S$ where $S$
has the form (\ref{ass}); this determines $X$ and $Y$ and hence a Gaussian
state $\psi_{X,Y}$ satisfying (\ref{wg}).

\section{Proof of Theorem \ref{th1}}

It is sufficient to assume that $\rho$ is centered at the origin, that is
$\langle z\rangle=0$.

\noindent\textbf{First step}. We begin by noting that the saturation of one of
the RS inequalities, for instance
\begin{equation}
(\Delta^{\rho}X_{1})^{2}(\Delta^{\rho}P_{1})^{2}=\Delta^{\rho}(X_{1}%
,P_{1})^{2}+\tfrac{1}{4}\hbar^{2},\label{1bis}%
\end{equation}
implies that $c(\Omega)=\frac{1}{2}h$. Let us prove this by \textit{reductio
ad absurdum}. Suppose that $c(\Omega)>\frac{1}{2}h$ and let $D=S^{T}MS$ be a
symplectic diagonalization (\ref{will}) of $M=\frac{\hbar}{2}\Sigma^{-}$.
Since $c(\Omega)$ is a symplectic invariant ($i.e.$ $c(S(\Omega))=c(\Omega)$
for every $S\in\operatorname*{Sp}(2n,\mathbb{R})$), our assumption can be
rewritten $c(\Delta)>\frac{1}{2}h$ where $\Delta$ is the ellipsoid defined by
$z^{T}Dz\leq\hbar$. In view of formula (\ref{CM}) we have $c(\Delta)=\pi
\hbar/\lambda_{1}$, recalling that the $\lambda_{j}$ form a decreasing
sequence. The assumption $c(\Delta)>\frac{1}{2}h$ is thus equivalent to
$\lambda_{1}<1$ and hence to $\lambda_{j}<1$ for all $j=1,...,n$. Consider now
the matrix%
\[
D^{-1}+iJ=%
\begin{pmatrix}
\Lambda^{-1} & I\\
-I & \Lambda^{-1}%
\end{pmatrix}
\geq0.
\]
Its eigenvalues are the roots of the polynomial
\[
P(t)=\prod\nolimits_{j=1}^{n}\left[  (\lambda_{j}^{-1}-t)^{2}-1\right]
\]
and are thus the real numbers $t_{j}=\lambda_{j}^{-1}\pm1$. Since $\lambda
_{j}<1$ for every $j$ this means that $t_{j}>0$ for every $j$ and hence
$D^{-1}+iJ>0$; returning to the covariance matrix, it follows that we also
have $\Sigma+\frac{i\hbar}{2}J>0$. But this leads to a contradiction: in view
of Sylvester's criterion for definite positiveness the condition $\Sigma
+\frac{i\hbar}{2}J>0$ implies the following condition on the minors of
$\Sigma+\frac{i\hbar}{2}J$:%

\[%
\begin{vmatrix}
(\Delta^{\rho}X_{j})^{2} & \Delta^{\rho}(X_{j},P_{j})+\frac{i\hbar}{2}\\
\Delta^{\rho}(X_{j},P_{j})-\frac{i\hbar}{2} & (\Delta^{\rho}P_{j})^{2}%
\end{vmatrix}
>0
\]
that is%
\[
(\Delta^{\rho}X_{j})^{2}(\Delta^{\rho}P_{j})^{2}>\Delta^{\rho}(X_{j}%
,P_{j})^{2}+\tfrac{1}{4}\hbar^{2}%
\]
for all $j=1,...,n$.

\noindent\textbf{Second step}. Let us show that if $\Omega$ is a phase space
ellipsoid such that $c(\Omega)=\frac{1}{2}h$, then $\Omega$ contains a unique
quantum blob (\textit{i.e.} a symplectic ball with radius $\sqrt{\hbar}$); we
do not assume explicitly that $\Omega$ is a covariance ellipsoid: the argument
is quite general. We begin by noting that the condition $c(\Omega)=\frac{1}%
{2}h$ implies, by definition of a symplectic capacity, that $\Omega$ contains
the image by some $S\in\operatorname*{Sp}(2n,\mathbb{R})$ of a ball
$B_{\sqrt{\hbar}}$ (and no symplectic ball with larger radius). This
symplectic ball can be explicitly constructed: assume that $\Omega$ is given
by $z^{T}Mz\leq\hbar$, $M>0$. In view of Williamson's diagonalization theorem,
there exists $S\in\operatorname*{Sp}(2n,\mathbb{R})$ such that%
\begin{equation}
S^{T}MS=D=%
\begin{pmatrix}
\Lambda & 0\\
0 & \Lambda
\end{pmatrix}
\text{ \ , \ }\Lambda=\operatorname*{Spec}\nolimits_{\sigma}(M)=(\lambda
_{1},...,\lambda_{n}). \label{sms}%
\end{equation}
(cf. (\ref{will}). The inequality $z^{T}Mz\leq\hbar$ is equivalent to%
\begin{equation}
\sum_{j=1}^{n}\lambda_{j}(x_{j}^{\prime2}+p_{j}^{\prime2})\leq\hbar\text{ \ ,
\ }(x^{\prime},p^{\prime})=S(x,p) \label{ellwi}%
\end{equation}
and we have $c(\Omega)=\pi\hbar/\lambda_{1}=\frac{1}{2}h$ so that $\lambda
_{1}=1$; it follows that the ellipsoid (\ref{ellwi}) contains $B_{\sqrt{\hbar
}}$, hence $\Omega$ contains the symplectic ball $S(B_{\sqrt{\hbar}})$. There
remains to prove the uniqueness of the a quantum blob contained in $\Omega$.
Using if necessary a phase space translation it is sufficient to consider the
case where $B_{\sqrt{\hbar}}:|z|\leq\hbar$ (in view of elementary geometric
considerations, the largest symplectic ball contained in $\Omega$ must be
centered at the origin: see \cite{Birk}, \S 8.4). Let us assume that there
exists $S^{\prime}\in\operatorname*{Sp}(2n,\mathbb{R})$ such that $S^{\prime
}(B_{\sqrt{\hbar}})$ is also contained in the ellipsoid $\Omega$; we are going
to show that $S^{\prime}(B_{\sqrt{\hbar}})=S(B_{\sqrt{\hbar}})$, following the
argument in de Gosson \cite{physlett,Bullsci,physreps}. Conjugating if
necessary $S$ and $S^{\prime}$ with an adequately chosen symplectic matrix we
may assume, using Williamson's theorem, that the $\lambda_{j}$ are the
symplectic eigenvalues of the matrix of $\Omega$. The condition $c(\Omega
)=\tfrac{1}{2}h$ then means that $\lambda_{1}=1$ in view of formula
(\ref{CM}). Let us show that $U=S^{\prime}S^{-1}$ belongs to the group of
symplectic rotations%
\begin{equation}
U(n)=\operatorname*{Sp}(2n,\mathbb{R})\cap O(2n,\mathbb{R}); \label{un}%
\end{equation}
the claim will follow since then $U(B_{\sqrt{\hbar}})=B_{\sqrt{\hbar}}$ so
that $S(B_{\sqrt{\hbar}})=S^{\prime}(B_{\sqrt{\hbar}})$. Clearly
$U\in\operatorname*{Sp}(2n,\mathbb{R})$ so it suffices to show that $U$ is in
addition a rotation; for this it is sufficient to check that $UJ=JU$. Set
$R=D^{1/2}UD^{-1/2}$ where $D=\operatorname*{diag}(\lambda_{1},...,\lambda
_{n})$; we have $U^{T}DU=D$ hence $R^{T}R=I$ so that $R$ is orthogonal. Let us
prove that $R$ is in addition symplectic. Since $J$ commutes with every power
of the diagonal matrix $D$ we have, taking into account the relation
$JU=(U^{T})^{-1}J$ (because $U$ is symplectic):%
\begin{align*}
JR  &  =D^{1/2}JUD^{-1/2}=D^{1/2}(U^{T})^{-1}JD^{-1/2}\\
&  =D^{1/2}(U^{T})^{-1}D^{-1/2}J=(R^{T})^{-1}J
\end{align*}
hence $R^{T}JR=J$ so that $R$ is in $\operatorname*{Sp}(2n,\mathbb{R})$. Since
$R$ is also a rotation we have $R\in U(n)$ and thus $JR=RJ$. Since
$U=D^{-1/2}RD^{1/2}$ we have%
\begin{align*}
JU  &  =JD^{-1/2}RD^{1/2}=D^{-1/2}JRD^{1/2}\\
&  =D^{-1/2}RJD^{1/2}=D^{-1/2}RD^{1/2}J\\
&  =UJ
\end{align*}
proving our claim.

\noindent\textbf{Third step}. We now define a normalized Gaussian state $\psi$
by specifying its Wigner transform:
\[
W\psi(z)=(\pi\hbar)^{-n}\exp\left[  -\frac{1}{\hbar}z^{T}(SS^{T}%
)^{-1}z\right]
\]
where $S\in\operatorname*{Sp}(2n,\mathbb{R})$ is defined as above,
\textit{i.e.} $S(B_{\sqrt{\hbar}})$ is the largest quantum blob contained in
$\Omega=\{z:z^{T}Mz\leq\hbar\}$. We note that the choice of $S$ is irrelevant,
because we have seen above that if $S(B_{\sqrt{\hbar}})=S^{\prime}%
(B_{\sqrt{\hbar}})$ then $S^{\prime}=SU$ with $U\in U(n)$ so that $S^{\prime
}S^{\prime T}=SS^{T}$. Since $S(B_{\sqrt{\hbar}})\subset\Omega$ we have
$z^{T}Mz\leq z^{T}(SS^{T})^{-1}z$ for all $z\in\mathbb{R}^{2n}$ and hence,
taking definition (\ref{rogauss}) of $W\rho(z)$ into account,
\begin{equation}
W\psi(z)\leq W\rho(z)\label{ouipsi2}%
\end{equation}
which is the inequality (\ref{ouipsi1}) in Theorem \ref{th1}. That $\psi$ is
the only Gaussian state satisfying this inequality follows at once from the
uniqueness of the quantum blob $S(B_{\sqrt{\hbar}})$. Let us finally show that
the states $\psi$ and $\rho$ have the same covariances $(\Delta^{\rho}%
X_{1})^{2}$, $(\Delta^{\rho}P_{1})^{2}$, and $\Delta^{\rho}(X_{1},P_{1})$. The
covariance matrix of $\psi$ is $\Sigma_{\psi}=\frac{\hbar}{2}SS^{T}$ and the
inclusion $S(B_{\sqrt{\hbar}})\subset\Omega$ implies that we have
$\Omega^{\ast}\subset S(B_{\sqrt{\hbar}})^{\ast}$ where $\Omega^{\ast}$ and
$S(B_{\sqrt{\hbar}})^{\ast}$ are the dual ellipsoids of $\Omega$ and
$S(B_{\sqrt{\hbar}})$, obtained by using a Legendre transformation (for a
review of the latter see \cite{zia}). Since the dual ellipsoid of an ellipsoid
$\frac{1}{2}z^{T}\Sigma z\leq1$ is $\frac{1}{2}z^{T}\Sigma^{-1}z\leq1$ this
implies that $S(B_{\sqrt{\hbar}})^{\ast}=(S^{T})^{-1}B_{2/\sqrt{\hbar}}$, so
the points of these dual ellipsoids satisfy
\begin{equation}
\frac{\hbar}{4}z^{T}SS^{T}z\leq\frac{1}{2}z^{T}\Sigma_{\rho}z\leq
1.\label{ellipsoids}%
\end{equation}
Let us cut $S(B_{\sqrt{\hbar}})^{\ast}=(S^{T})^{-1}B_{2/\sqrt{\hbar}}$ and
$\Omega^{\ast}=\{z:\tfrac{1}{2}z^{T}\Sigma_{\rho}z\leq1\}$ by the plane
$P_{1,1}$ of coordinates $x_{1},p_{1}$. This section consists of two ellipses:
$P_{1,1}\cap(S^{T})^{-1}B_{2/\sqrt{\hbar}}$, which has area $4\pi/\hbar$, and
\[
P_{1,1}\cap\Omega^{\ast}:\frac{1}{2}%
\begin{pmatrix}
x_{1} & p_{1}%
\end{pmatrix}%
\begin{pmatrix}
(\Delta^{\rho}X_{1})^{2} & \Delta^{\rho}(X_{1},P_{1})\\
\Delta^{\rho}(X_{1},P_{1}) & (\Delta^{\rho}P_{1})^{2}%
\end{pmatrix}%
\begin{pmatrix}
x_{1}\\
p_{1}%
\end{pmatrix}
\leq1;
\]
taking the saturation of the first RS inequality (\ref{1}) into account,
$P_{1,1}\cap\Omega^{\ast}$ also has area
\[
2\pi\left[  (\Delta^{\rho}X_{1})^{2}(\Delta^{\rho}P_{1})^{2}-\Delta^{\rho
}(X_{1},P_{1})^{2}\right]  ^{-12}=4\pi/\hbar.
\]
Since $P_{1,1}\cap\Omega^{\ast}\subset P_{1,1}\cap S(B_{\sqrt{\hbar}})^{\ast}$
equality of areas implies that the ellipses $P_{1,1}\cap\Omega^{\ast}$ and
$P_{1,1}\cap S(B_{\sqrt{\hbar}})^{\ast}$ are identical, hence the covariances
$(\Delta^{\rho}X_{1})^{2}$, $(\Delta^{\rho}P_{1})^{2}$, and $\Delta^{\rho
}(X_{1},P_{1})$ are the same for both states $\psi$ and $\rho$. \textbf{QED}.

\section*{Appendix: The Linear Gromov Theorem}

Let us prove the following linear version of Gromov's non-squeezing theorem.
We are following de Gosson \cite{Birkbis}, \S 5.1.2; we give two independent proofs.

\begin{proposition}
\label{proplingro}Let $S\in\operatorname*{Sp}(2n,\mathbb{R})$ and
$B_{R}:|z|\leq R$ The intersection of $S(B_{R})$ by a plane of conjugate
coordinates $x_{j},p_{j}$ is an ellipse with area $\pi R^{2}$.
\end{proposition}

\noindent\textbf{First proof}\textit{. }It relies on the fact that the form
$pdx=\sum_{j}p_{j}dx_{j}$ is a relative symplectic integral invariant, that
is: if $\phi$ is a symplectomorphism of $\mathbb{R}^{2n}$ and $\gamma$ a loop
in $\mathbb{R}^{2n}$, then%
\begin{equation}
\oint\nolimits_{\gamma}pdx=\oint\nolimits_{\phi(\gamma)}pdx \label{pdx}%
\end{equation}
(see for instance Arnol'd \cite{Arnold}, \S 44, p.239). We claim that the
ellipse $\Gamma_{j}=S(B_{R})\cap\mathcal{P}_{j}$, intersection of the
ellipsoid $S(B_{R})$ with any plane $\mathcal{P}_{j}$ of conjugate coordinates
$x_{j},p_{j}$ has area $\pi R^{2}$; the proposition immediately follows from
this property. Let $\gamma_{j}$ be the curve bounding the ellipse $\Gamma_{j}$
and orient it positively; the area it encloses is
\begin{equation}
\operatorname*{Area}(\Gamma_{j})=\oint\nolimits_{\gamma_{j}}pdx=\oint%
\nolimits_{S^{-1}(\gamma_{j})}pdx=\pi R^{2} \label{gj}%
\end{equation}
(because $S^{-1}(\gamma_{j})$ is a big circle of $B_{R}$); notice that the
assumption that $\mathcal{P}_{j}$ is a plane of conjugate coordinates
$x_{j},p_{j}$ is essential for (\ref{gj}) to hold, making the use of the
formula (\ref{pdx}) possible [more generally, the argument works when
$\mathcal{P}_{j}$ is replaced by any symplectic plane].

\noindent\textbf{Second proof}\textit{. }With the same notation as above we
note that the set
\[
S^{-1}\left[  S(B_{R})\cap\mathcal{P}_{j}\right]
\]
is a big circle of $B_{R}$, and hence encloses a surface with area $\pi R^{2}%
$. Now, $\mathcal{P}_{j}$ is a symplectic space when equipped with the
two-form $\sigma_{j}=dp_{j}\wedge dx_{j}$ and the restriction of $S$ to
$\mathcal{P}_{j}$ is a linear symplectomorphism from $(\mathcal{P}_{j}%
,\sigma_{j})$ to the symplectic plane $S(\mathcal{P}_{j})$ equipped with the
restriction of the symplectic form $\sigma$. Symplectomorphisms being volume
(here: area) preserving, it follows that $S(B_{R})\cap\mathcal{P}_{j}$ also
has area $\pi R^{2}$.\bigskip

\noindent\textbf{Remark.} It would certainly be interesting to generalize the
first proof to arbitrary symplectomorphisms. The difficulty comes from the
following fact: the key to the proof in the linear case is the fact that we
were able to derive the equality%
\[
\int_{\gamma_{R}}p_{j}dx_{j}=\pi R^{2}%
\]
by exploiting the fact the inverse image of the $x_{j},p_{j}$ plane by $S$ was
a plane cutting $B_{R}$ along a big circle, which thus encloses an area equal
to $\pi R^{2}$. When one replaces the \emph{linear} transformation $S$ by a
non-linear one, the inverse image of the $x_{j},p_{j}$ plane will not
generally be a plane, but rather a 2-dimensional symplectic manifold. If the
following property holds: \emph{The section of }$B_{R}$\emph{\ by any
2-dimensional symplectic manifold containing the center of }$B_{R}$\emph{\ has
an area at least }$\pi R^{2}$ then we would have, by the same argument as
above%
\[
\int_{\gamma_{R}}p_{j}dx_{j}\geq\pi R^{2}.
\]
We do not know any proof of this property; nor do we know whether it is true!

\end{document}